\documentclass[twocolumn,preprintnumbers,aps,prb,amsmath,amssymb,floatfix]{revtex4}
\usepackage{graphicx}
\usepackage{bm}

\begin{document}

\title{Investigation of the crystal and magnetic structures of the trigonal multiferroic iron-boracite Fe$_3$B$_{7}$O$_{13}$(OH)}

\author{Gwilherm N\'enert$^{1}$\footnote{Corresponding author: Gwilherm N\'{e}nert, g.nenert@gmail.com}, Ippei Nomoto$^2$, Hirohiko Sato$^2$}

\address{$^{1}$ Institut Laue-Langevin, Bo\^{i}te Postale 156, 38042 Grenoble Cedex 9 France}
\address{$^{2}$ Department of Physics, Chuo University, 1-13-27 Kasuga, Bunkyo, Tokyo 112-8551, Japan}

(Dated: \today)

\begin{abstract}
We have investigated the crystal and magnetic structures of the trigonal iron-boracite Fe$_3$B$_{7}$O$_{13}$X with X = OH by neutron diffraction. Neutron diffraction enables us to locate the hydrogen atom of the hydroxyl group and determine the magnetic ground state of this member of the multiferroic boracite family. No evidence was found for a monoclinic distortion in the magnetic ordered state. The magnetic symmetry allows for magnetoelectric and ferroelectric properties. The N\'eel temperature T$_N$ of 4.86(4) K confirms the general trends within the boracites that T$_N$ decreases from X = I $>$ Br $>$ Cl $>$ OH. Surprisingly while Fe$_3$B$_{7}$O$_{13}$OH exhibits the largest frustration with $\mid\theta/T_N\mid$ = 5.6 within the Fe$_3$B$_{7}$O$_{13}$X series, no reduction of the magnetic moment is found using neutron diffraction.

\end{abstract}
\pacs{PACS numbers: }

\maketitle

\section{Introduction}

Since the discovery of multiferroic properties in TbMnO$_3$,\cite{Kimura} the field of multiferroic 
materials have attracted a lot of interest.\cite{Fiebig,Eerenstein,Maxim} This interest arises from the emergence of 
new fundamental physics\cite{electromagnon} and potential
technological applications.\cite{Eerenstein,Maxim} This field also gave rise to the reinvestigation of a large number 
of "old" materials such as, for instance, the manganites RMnO$_3$ \cite{RMnO3} or the pyroxene family.\cite{pyroxene} Historically, the first family of multiferroic materials to be investigated was the boracite family. 

Boracites are materials exhibiting the general formula M$_3$B$_7$O$_{13}$X where M is a transition metal ion 
or alternatively Mg, Cd. The vast majority of the boracites are halogen boracites with compositions X = Cl, Br or I.\cite{boracites} Occasionally X can be OH, F or NO$_3$ and these associated phases have been much less investigated.\cite{M3B7O13X} The boracites have been widely investigated due to their ferroelectric, ferroelastic and magnetic properties.\cite{multiferroic} Several compositions within the boracites with X = Cl are natural minerals. They are of interest for mineralogists due their complex twinning and anomalous optical properties.\cite{minerals} Despite the large number of studies dedicated to this family and its wide chemistry, few studies have been dedicated to the determination of their magnetic ground states using neutron diffraction.\cite{Mn3B7O13I,Ni3B7O13Br,Co3B7O13Br,Ni3B7O13Cl}

Recently, a new composition with M = Fe and X = OH has been reported.\cite{Fe3B7O13OH} It has been shown that this boracite crystallizes in the space group \textit{R3c} (No. 161). This system orders antiferromagnetically below T$_N$ $\simeq$ 4.8 K and potentially exhibits magnetic frustration. Magnetic frustration could arise due to the arrangement of magnetic Fe$^{2+}$ ions which is based on a triangular framework. A magnetic system is considered to be spin frustrated when the
ratio f = $\mid\theta/T_N\mid$ is equal to or greater than 6.\cite{frustration} For Fe$_3$B$_7$O$_{13}$OH, f
is about 5.6,\cite{Fe3B7O13OH} and thus it may exhibit some magnetic frustration. However in the absence of neutron diffraction, this study could not further probe the exact nature of the ground state of this material. We aim here to investigate the magnetic ground state using powder and single crystal neutron diffraction. Additionally, we have used neutron diffraction in order to better characterize the crystal structure and in particular the hydrogen position which could not be located from x-ray single crystal work.

\section{Experiment}

Small single crystals of Fe$_3^{11}$B$_{7}$O$_{13}$(OH) were synthesized
by a hydrothermal method. A mixture of FeO, $^{11}$B$_{2}$O$_3$, and NaOH solution (4 mol/L) was
sealed in a silver capsule. Then it was heated up to 600 $^{\circ}$C
in a test-tube-type autoclave under 150 MPa of hydrostatic
pressure. After the reaction for 3 days, the product was
washed with hot water in order to remove the excess of $^{11}$B$_{2}$O$_3$. $^{11}$B$_{2}^{11}$O$_3$ was used in order to reduce the absorption of natural boron by neutrons.

Most of the neutron diffraction measurements were carried out on powder
samples. The precise crystal and magnetic structures
were investigated using high resolution powder data at various temperatures 
using the D2B diffractometer at the Institut Laue Langevin (ILL). The measurements were
carried out at a wavelength of 1.594 \r{A} corresponding to the
(335) Bragg reflexion of a germanium monochromator. The neutron
detection is performed with $^{3}$He counting tubes spaced at
1.25$^{\circ}$ intervals for D2B. A complete diffraction pattern
is obtained after about 25 steps of 0.05$^{\circ}$ in 2$\theta$.

Powder neutron diffraction was carried out by crushing small single crystals resulting in a fine light brown powder. Measurement was carried out above the N\'eel temperature (T $\sim$ 9 K) and below (T = 1.8 K). Diffraction data analysis was done using the FullProf refinement package. \cite{fullprof}

Additional data were collected on the high resolution four-circle single crystal diffractometer D9 at the ILL. Few reflections were followed as function of temperature to determine the critical temperature behavior of the magnetic order. Data collection was done using a wavelength of 0.706 \r{A} obtained by reflection from a Cu(220) monochromator. The wavelength was calibrated using a germanium single crystal. D9 is equipped with a small two- dimensional area detector\cite{lehmann}, which for this measurement allowed optimal delineation of the peak from the background. For all data, background corrections\cite{Wilkinson} and Lorentz corrections were applied. 

\section{Results and discussion}

\subsection{Structural properties}\label{Structural properties}

Attempts to solve the crystal structure using single crystal neutron diffraction data were unsuccessful due to large twinning of the crystals. Consequently only few reflections measured on the single crystal diffractometer could be used. Powder diffraction data were used to solve the crystal and magnetic structures.

\begin{figure}[htb]
\centering
\includegraphics[angle=-90,width=8cm]{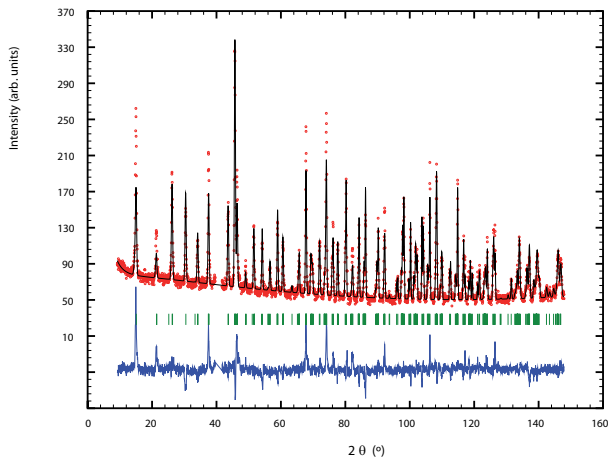}
\caption{(Color Online) Refinement of neutron data at 9 K of the crystal structure of Fe$_3$B$_{7}$O$_{13}$(OH) using the structural model derived from single crystal data. The excluded region around 40 degrees is to remove the cryostat contribution.}\label{Cell9K_No-H}
\end{figure}

Attempts to refine the crystal structure at 9 K using the x-ray single crystal model were unsuccessful. The best refinement which could be obtained is shown in Figure \ref{Cell9K_No-H}. These data show clearly that some intensity is lacking over the whole pattern. This discrepancy results from the impossibility from the x-ray single crystal data to locate the hydrogen atom of the hydroxyl group. Prior to investigate the magnetic properties of this polar iron boracite, we have been using the neutron diffraction data in order to locate the hydrogen atom of the hydroxyl group.

\begin{figure}[htb]
\centering
\includegraphics[width=8cm]{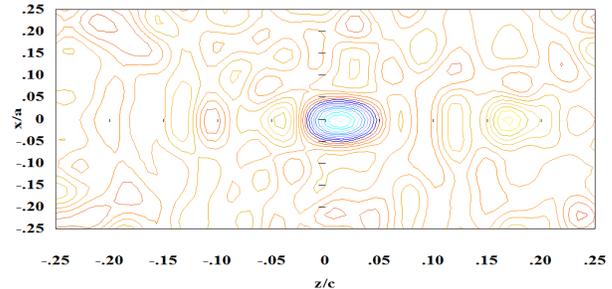}
\caption{(Color Online) Difference Fourier map showing the presence of the hydrogen atom sitting on the Wyckoff position 6a (0 0 z $\sim$ 0.03).}\label{FourierMap}
\end{figure}

\begin{figure}[htb]
\centering
\includegraphics[angle=-90,width=8cm]{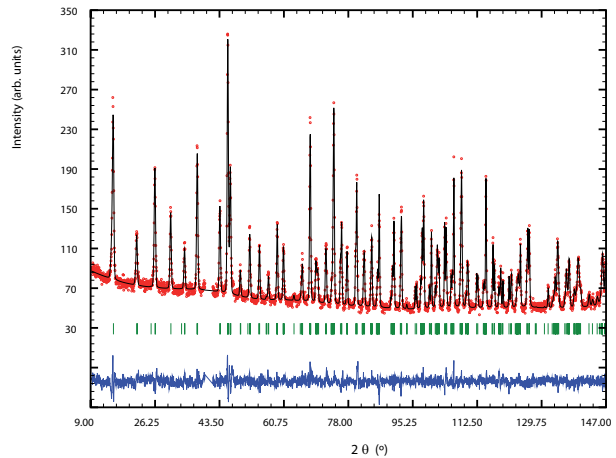}
\caption{(Color Online) Refinement of neutron data at 9 K of the crystal structure of Fe$_3$B$_{7}$O$_{13}$(OH) including the hydrogen atom of the hydroxyl group in (0, 0, z = 0.0426(10)). The excluded region around 40 degrees is to remove the cryostat contribution. Statistics: R$_p$=3.34\% and R$_{Bragg}$ = 5.57\%}\label{9K_H}
\end{figure}

\begin{figure}[htb]
\centering
\includegraphics[width=3cm]{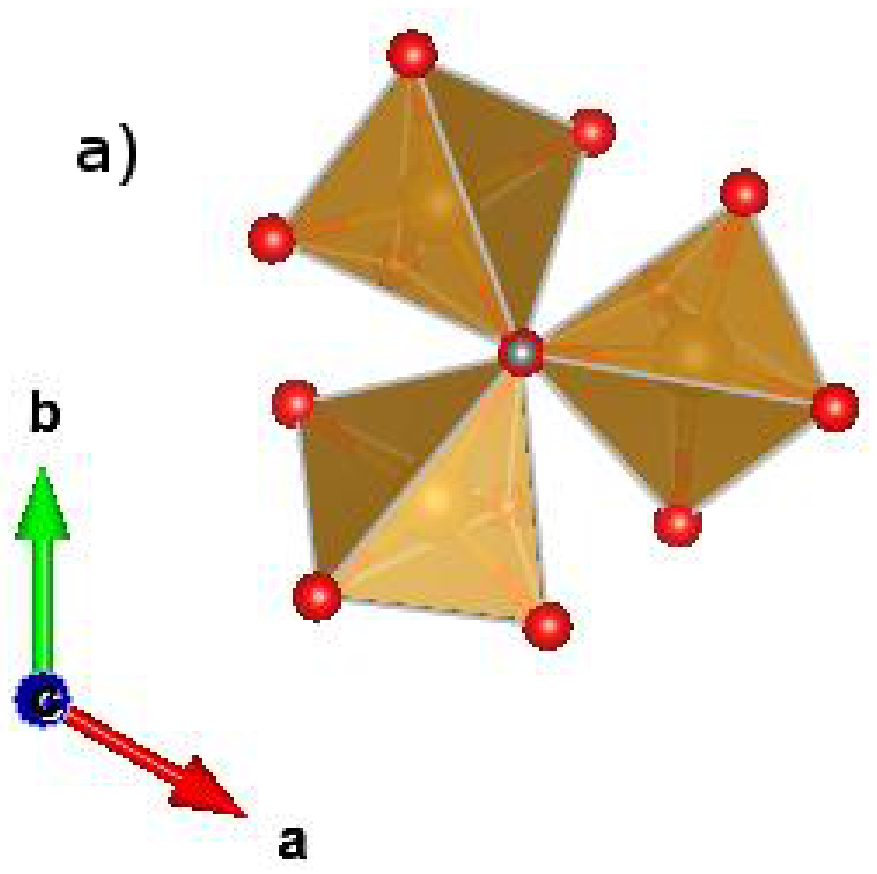}
\includegraphics[width=3cm]{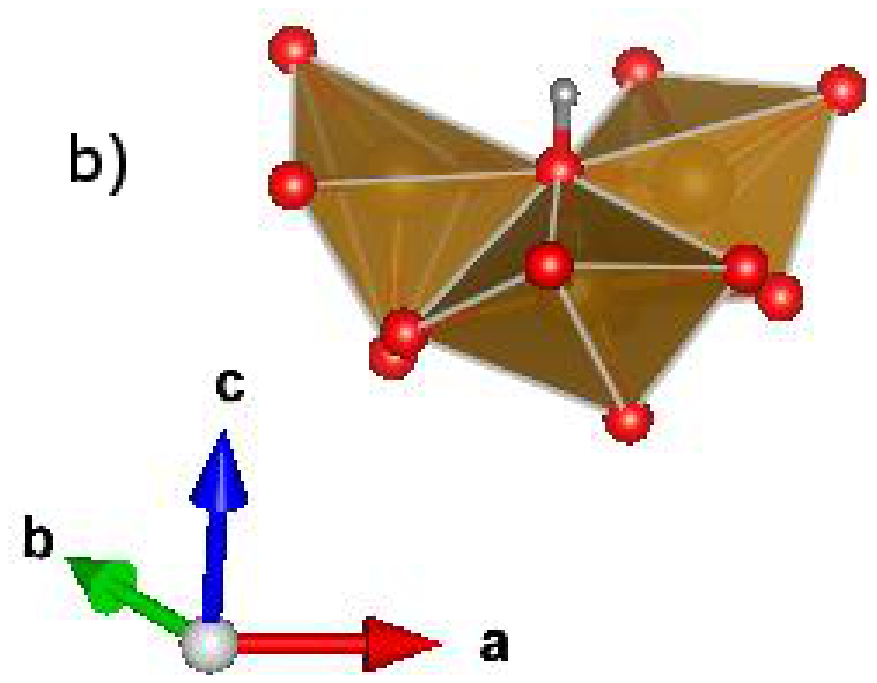}\\
\caption{(Color Online) a) Detail of the Fe$_3$ trimer unit. The center oxygen is actually a OH$^{-}$ ion which is illustrated in b). The O - H bond is directed along the \textit{c} axis. Drawing was made using the software VESTA.\cite{VESTA}}\label{Iron_Polyhedra}
\end{figure}

Localization of the missing hydrogen atom participating to the hydroxyl group was done by calculating the difference Fourier map of the refined pattern shown in Figure \ref{Cell9K_No-H}. The difference Fourier map obtained at 9 K is illustrated in Figure \ref{FourierMap}. The hydrogen atom can be localized in the Wyckoff position \textit{6a} in (0, 0, z). The refined atomic position of the hydrogen atom is (0, 0, z = 0.0426(10). The final Rietveld refinement at 9 K is show in Figure \ref{9K_H} and the corresponding atomic positions are given in Table \ref{structure10K}. A representation of the Fe$_3$ trimer unit with the hydroxyl group is shown in Figure \ref{Iron_Polyhedra}. The O - H bond is directed along the polar \textit{c} axis. Its distance obtained after refinement is  1.00(3) \r{A}. This bond distance is in excellent agreement with other report for hydroxyl group in minerals.\cite{eosphorite}

\begin{table}[htb]
\centering
\begin{tabular}{c c c c c c}
\hline \hline
Atom & Wyckoff &x& y& z& U$_{iso}$ \\
\hline
Fe    & 18b & 0.5247(7) &   1.0580(4) &   0.2993(5) &   0.0057(4)\\ 
O$_1$ &  6a & 0.66666   &   0.33333   &   0.32818(-)   &   0.0137(20)\\     
H     &  6a & 0.00000   &   0.00000   &   0.0426(10)&   0.017(3)\\  
O$_2$ & 18b & 0.7085(9) &   0.9748(10)&   0.3254(6) &   0.0094(13)\\
O$_3$ & 18b & 0.6445(8) &   0.1092(9) &   0.2097(5) &   0.0068(12)\\
O$_4$ &  6a & 0.00000   &   0.00000   &   0.2999(7) &   0.0047(16)\\   
O$_5$ & 18b & 0.3397(9) &   0.8310(9) &   0.3502(6) &   0.0082(12)\\
O$_6$ & 18b & 0.3043(10)&   0.0814(9) &   0.2676(6) &   0.0073(11)\\
B$_1$ & 18b & 0.1740(9) &   0.8349(8) &   0.3712(5) &   0.0080(9)\\ 
B$_2$ &  6a & 0.33333   &   0.66666   &   0.3524(6) &   0.0049(15)\\          
B$_3$ & 18b & 0.8973(10)&   0.0996(9) &   0.3165(6) &   0.0070(8)\\ 
\hline \hline
\end{tabular}
\\
\caption{Crystallographic coordinates extracted from the Rietveld
refinement carried out on powder neutron diffraction (D2B) using
the space group \textit{R3c} at 9 K with cell parameters \emph{a}
= \emph{b} = 8.56080(5) \r{A}  and \emph{c} =
21.06236(19) \r{A}. The z coordinate of the O$_1$ has been fixed in order to define the origin.}\label{structure10K}
\end{table}

\subsection{Magnetic structure}\label{Magnetic structure}

\begin{figure}[htb]
\centering
\includegraphics[angle=-90,width=8cm]{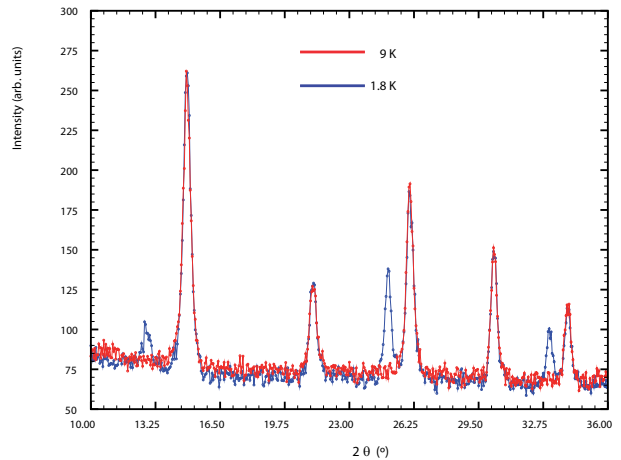}
\caption{(Color Online) Powder diffraction patterns recorded at 9 and 2 K, shown respectively in red and blue. All the magnetic reflections can be indexed on the chemical unit cell.}\label{Difference}
\end{figure}

As mentioned in the previous section, the magnetic structure was determined using the powder neutron diffraction data. The 1.8 K neutron diffraction pattern collected on D2B indicates the presence of additional magnetic reflections at reciprocal lattice positions of the nuclear cell as shown in Figure \ref{Difference}.

\begin{figure}[htb]
\centering
\includegraphics[width=8cm]{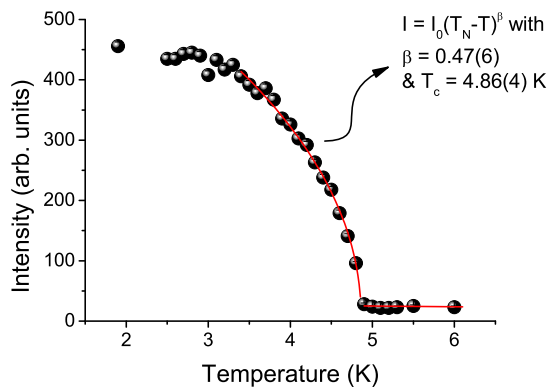}
\caption{(Color Online) Integrated intensity of the (205) magnetic reflection as a
function of temperature. The line corresponds to a fit to the power law
I = I$_0$(T$_N$-T)$^\beta$ in the vicinity of T$_N$ = 4.86(4) K and constant above.}\label{205_vs_T}
\end{figure}

Using single crystal data, despite of the twinning, we could follow as function of temperature few magnetic reflections. This enables us to probe the nature of the magnetic phase transition. We present in Figure \ref{205_vs_T} the temperature evolution of the (205) reflection. Attempt to fit the data close to T$_N$ using a phenomenological model of power law gives rise to T$_N$ = 4.86(4) K. This N\'eel temperature is in excellent agreement with the previously reported value.\cite{Fe3B7O13OH} The critical exponent that we obtain give rise to $\beta$ = 0.47(6) which is close to $\frac{1}{2}$ suggesting that the magnetic ordering in Fe$_3$B$_{7}$O$_{13}$(OH) follows a typical mean-field theory.

The possible magnetic structures compatible with the
symmetry of Fe$_3$B$_{7}$O$_{13}$(OH) were determined using BasIreps.\cite{BasiReps} For
the propagation vector $\overrightarrow{k}$ = $\overrightarrow{0}$, the small group G$_{\overrightarrow{k}}$, formed by
those elements of the space group that leave $\overrightarrow{k}$ invariant,
coincides with the space group \textit{R3c}. For $\overrightarrow{k}$ = $\overrightarrow{0}$, the irreducible
representations of the group G$_{\overrightarrow{k}}$ are those shown in Table
\ref{irreps}.

\begin{table*}[htb]
\centering
\caption{Irreducible representations of the space group \textit{R3c} for $\protect\overrightarrow{k}$=$\protect\overrightarrow{0}$. The symmetry elements are written according to Seitz notation (Ref. \cite{Seitz})}
\begin{tabular}{c c c c c c c}
\hline \hline
 & 1$\vert$0,0,0 & {3+$_{00z}\vert$000} & {3-$_{00z}\vert$000}  & (m$_{x\overline{x}z}\vert$0,0,$\frac{1}{2}$) & (m$_{x2xz}\vert$0,0,$\frac{1}{2}$) & (m$_{2xxz}\vert$0,0,$\frac{1}{2}$)\\
\hline
 $\Gamma_{1}$     &  1 &  1  &  1 &  1 &  1 &  1\\
 $\Gamma_{2}$     &  1 &  1  &  1 & -1 & -1 & -1\\
 $\Gamma_{3}$     &  $\begin{pmatrix}
 1&0\\
 0&1
\end{pmatrix}$  & $\begin{pmatrix}
 -\frac{1}{2}+\frac{\sqrt{3}}{2}i&0\\
 0&-\frac{1}{2}-\frac{\sqrt{3}}{2}i
\end{pmatrix}$ & $\begin{pmatrix}
 -\frac{1}{2}-\frac{\sqrt{3}}{2}i&0\\
 0&-\frac{1}{2}+\frac{\sqrt{3}}{2}i
\end{pmatrix}$& $\begin{pmatrix}
 0&1\\
 1&0
\end{pmatrix}$ & $\begin{pmatrix}
0 & -\frac{1}{2}-\frac{\sqrt{3}}{2}i\\
 -\frac{1}{2}+\frac{\sqrt{3}}{2}i & 0
\end{pmatrix}$ & $\begin{pmatrix}
0 & -\frac{1}{2}+\frac{\sqrt{3}}{2}i\\
 -\frac{1}{2}-\frac{\sqrt{3}}{2}i & 0
\end{pmatrix}$\\
\hline \hline
\end{tabular}
\label{irreps}
\end{table*}

A representation $\Gamma$ is constructed with the Fourier components
\textbf{m$^k$} corresponding to the Fe atoms of the Wyckoff
position 18b. The decomposition of the representation 
$\Gamma$ in terms of the irreducible representations $\Gamma_{\overrightarrow{k}}$ is for the Wyckoff 18b site,

\begin{equation}
\Gamma_{\overrightarrow{k}}(18b) = \Gamma_1 + \Gamma_2 + 2\Gamma_3
\end{equation}

\begin{figure}[htb]
\centering
\includegraphics[angle=-90,width=8cm]{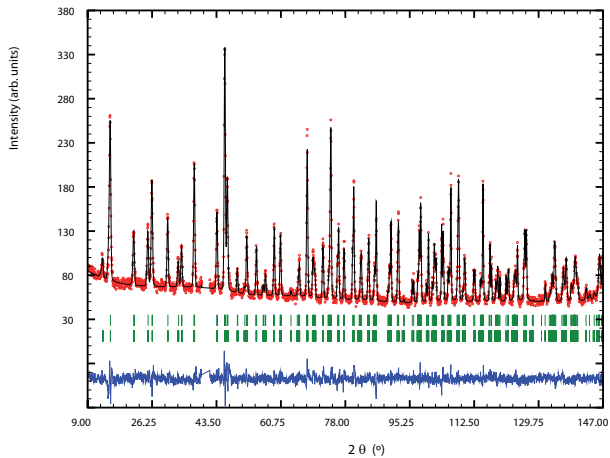}
\caption{(Color Online) Refinement of neutron data at 1.8 K of the magnetic structure of Fe$_3$B$_{7}$O$_{13}$(OH). The excluded region around 40 degrees is to remove the cryostat contribution. Statistics: R$_p$=3.27\%, R$_{Bragg}$ = 5.86\% and R$_{mag}$ = 7.6\%.}\label{Rietveld_Mag}
\end{figure}

\begin{figure}[htb]
\centering
\includegraphics[width=6cm]{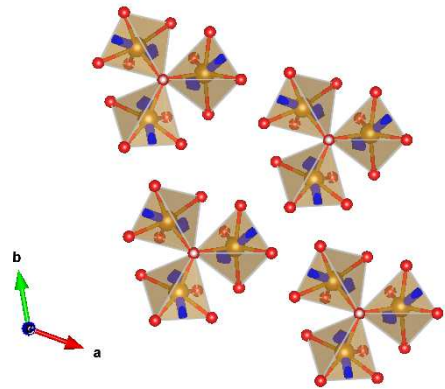}
\caption{(Color Online) Representation of the magnetic structure obtained at 1.8 K within the \textit{ac} plane. For the next layer along the \textit{c} axis, the spins are rotated by 60$^{\circ}$. Graphical representation was made using the software VESTA.\cite{VESTA}}\label{MagneticStructure}
\end{figure}

The best refinement of the powder neutron data was obtained considering the magnetic structure associated to the irreducible representation $\Gamma_1$. The resulting magnetic moment for the Fe$^{2+}$ ions is 4.5(2) $\mu_B$. This value is higher than the spin only value of 4 $\mu_B$. This is likely related to the orbital contribution to the magnetic moment coming from Fe$^{2+}$ ions. This is in agreement with the reported M\"{o}ssbauer data where a large quadrupole spitting is reported.\cite{Fe3B7O13OH} The resulting fit of the powder data at 1.8 K is presented in Figure \ref{Rietveld_Mag}. A representation of the magnetic structure is shown in Figure \ref{MagneticStructure}. The spins lie mostly within the (ab) plane with a small out of plane component. This out of plane component is necessary in order to describe properly the first magnetic reflection at 2$\theta\sim$13.1$^{\circ}$. In Figure \ref{MagneticStructure}, the out of plane component points down. While moving along the trigonal axis, the orientation of the spins changes by 60$^{\circ}$ and the out of plane component points alternatively down and up. The overall magnetic structure is purely antiferromagnetic without any weak ferromagnetic component. The resulting magnetic symmetry is \textit{R3c}.

These results are in contrast with the reported literature on the magnetic ground state of the boracites.\cite{boracites,Mn3B7O13I,Ni3B7O13Br,Co3B7O13Br,Ni3B7O13Cl} Fe$_3$B$_{7}$O$_{13}$(OH) crystallizes in the trigonal space group \textit{R3c} which is potentially ferroelectric but does not exhibit any weak ferromagnetic component in contrast to all the other reported boracites.\cite{boracites} Even Co$_3$B$_7$O$_{13}$Cl which also exhibits the trigonal symmetry \textit{R3c} at room and low temperature, changes of symmetry giving rise to a weak ferromagnetic component below T$_N$ = 12 K.\cite{Co3B7O13Cl} It would be of interest to investigate the other compositions exhibit the trigonal \textit{R3c} symmetry (X = OH, NO$_3$ for instance) at room temperature in order to investigate further whether Fe$_3$B$_{7}$O$_{13}$(OH) is the exception to the rule or not.

Another interesting point in the magnetic properties of Fe$_3$B$_{7}$O$_{13}$(OH) is the absence of reduction of the magnetic moment despite the expected presence of magnetic frustration (f = 5.6). While all the investigated compositions by neutron diffraction exhibit the same magnetic symmetry Pc$^{'}$a2$_1^{'}$ and a lower symmetry than Fe$_3$B$_{7}$O$_{13}$(OH), their resulting magnetic moments are lower than the spin only values. In the paramagnetic space group Pca2$_1$1$^{'}$, there are 3 different crystallographic site for M. Function of the metal M, the neutron experiments show that there is one or two crystallographic sites exhibiting magnetic frustration giving rise to a reduced magnetic moment. For M = Co and X = Br, despite the presence of the expected large orbital momentum contribution, the third site exhibits a magnetic moment of 1.7(1) $\mu_B$ while the other 2 show respectively a magnetic moment of 4.7(2) and 4.0(2) $\mu_B$.\cite{Co3B7O13Br} For Mn$_3$B$_{7}$O$_{13}$I, only one site shows a saturated magnetic moment with 5.4(2) $\mu_B$ while the other two sites are reported with a magnetic moment of 3.8(2) $\mu_B$.\cite{Mn3B7O13I} Similar results are reported for the other compositions investigated by neutron diffraction.\cite{Ni3B7O13Br,Ni3B7O13Cl} In boracites, the frustration parameter f increases going from X = I $>$ Br $>$ Cl.\cite{Mn3B7O13I,Fe3B7O13X} Using the results from literature for M = Fe, we can further extend the rule to X = OH and we notice that the frustration parameter f increases from X = OH $>$ I $>$ Br $>$ Cl going from 1.4 (X = Cl)  to 5.6 ( X = OH). The f parameter for the halogen boracites remains small irrespective of the chemical composition and much below 6 (at most f $\sim$ 3 for X = I). Consequently the lowering of symmetry from \textit{R3c} to \textit{Pca2$_1$} gives rise surprisingly to an increase of the reduction of the magnetic moment while the frustration parameter f decreases. DFT calculations would be necessary in order to investigate in more detail the magnetic frustration in the boracites.

\section{Conclusion}

We have investigated by neutron diffraction the crystal and magnetic structures of the newly reported trigonal iron boracite Fe$_3$B$_{7}$O$_{13}$(OH). We were able to locate the hydrogen atom within the structure by Fourier map difference. The hydroxyl group is characterized by a hydrogen oxygen bond distance in excellent agreement with other hydroxyl groups reported in other minerals. In agreement with previous report, we find that below T$_N$ = 4.86(4) K an antiferromagnetic state takes place characterized by $\overrightarrow{k}$ = $\overrightarrow{0}$. The resulting magnetic moment is 4.5(2) $\mu_B$ is larger than the spin only value of 4 $\mu_B$. This difference is probably related to the orbital contribution to the magnetic moment. We show that the magnetic frustration in boracites increases along X = OH $>$ I $>$ Br $>$ Cl although without giving rise to a reduced magnetic moment for X = OH as expected for a magnetically frustrated system. We demonstrate that Fe$_3$B$_{7}$O$_{13}$(OH) is a very unusual system within the boracite family. We expect that this work will stimulate experimental investigations of the other compositions with the boracite family with X = OH and NO$_3$.

\section*{ACKNOWLEDGEMENTS}
The authors acknowledge the allocation of beamtime at the Institut Laue Langevin and the technical support provided during the experiment.


\begin{references}

\bibitem{Kimura} T. Kimura, T. Goto, H. Shintani, K. Ishizaka, T. Arima, Y. Tokura, Nature \textbf{426}, 55-58 (2003).\label{Kimura}
\bibitem{Fiebig} M. Fiebig, J. Phys. D: Appl. Phys. \textbf{38} R123 (2005).\label{Fiebig}
\bibitem{Eerenstein} W. Eerenstein, N. D. Mathur and J. F. Scott, Nature \textbf{442} 759 (2006).\label{Eerenstein}
\bibitem{Maxim} S.-W. Cheong and M. Mostovoy, Nature Materials \textbf{6}, 13 (2007).\label{Maxim}
\bibitem{electromagnon} A. Pimenov, A. A. Mukhin, V. Yu. Ivanov, V. D. Travkin, A. M. Balbashov, A. Loidl; Nature Physics \textbf{2}, 97 (2006); A. B. Sushkov, R. Vald\'{e}s Aguilar, S. Park, S-W. Cheong, H. D. Drew; Phys. Rev. Lett. \textbf{98}, 027202 (2007).\label{electromagnon}
\bibitem{RMnO3} N. Aliouane, O. Prokhnenko, R. Feyerherm, M. Mostovoy, J. Strempfer, K. Habicht, K. C. Rule, E. Dudzik, A. U. B. Wolter, A. Maljuk, D. N. Argyriou, J. Phys.: Condens. Matter 20 434215 (2008); B. Lorenz, ISRN Condensed Matter Physics, vol. 2013, Article ID 497073, 43 pages, 2013. doi:10.1155/2013/497073. And references therein.\label{RMnO3}
\bibitem{pyroxene} S. Jodlauk, P. Becker, J. A. Mydosh, D. I. Khomskii, T. Lorenz, S. V. Streltsov, D. C. Hezel and L. Bohat\'{y}, J. Phys.: Condens. Matter \textbf{19} 432201 (2007); G. N\'{e}nert, M. Isobe, C. Ritter, O. Isnard, A. N. Vasiliev, Y. Ueda, Phys. Rev. B \textbf{79}, 064416 (2009); G. N\'{e}nert, I. Kim, M. Isobe, C. Ritter, A. N. Vasiliev, K. H. Kim, Y. Ueda, Phys. Rev. B \textbf{81}, 184408 (2010); G. N\'{e}nert, M. Isobe, I. Kim, C. Ritter, C. V. Colin, A. N. Vasiliev, K. H. Kim, Y. Ueda, Phys. Rev. B. \textbf{82}, 024429 (2010).\label{pyroxene}
\bibitem{boracites} R. Nelmes; J. Phys. C: Solid State Phys., Vol. \textbf{7}, 3840-3854 (1974).\label{boracites}
\bibitem{M3B7O13X} T. A. Bither, H. S. Young; Journal of Solid State Chemistry \textbf{10}, 302–311 (1974).\label{M3B7O13X}
\bibitem{multiferroic} See for instance: J.-P. Rivera, H. Schmid, J. M. Moret, H. Bill, International Journal of Magnetism, vol. \textbf{6}, no. 3-4, p. 211-220 (1974); J. Kacz\'{e}r, T. Shalnikova, Z. Hauptman, J. Appl. Phys. \textbf{39}, 429 (1968); E. Ascher, H. Rieder, H. Schmid, H. St\"{o}ssel, J. Appl. Phys. \textbf{37}, 1404-1405 (1966); O. Crottaz, J.-P. Rivera, B. Revaz, H. Schmid, Ferroelectrics, \textbf{204} 125-133 (1997).\label{multiferroic}
\bibitem{minerals} P. C. Burns, Powder Diffraction \textbf{10}, 250-260 (1995); P. C. Burns, M. A. Carpenter; Can. Mineral. \textbf{34}, 881-892 (1996); P. C. Burns, J. D. Grice, F. C. Hawthorne, Can. Mineral. \textbf{33}, 1131-1151 (1995); P. C. Burns, F. C. Hawthorne, J. A. R. Stirling, Can. Mineral. \textbf{30}, 445-448 (1992).\label{minerals}
\bibitem{Mn3B7O13I} O. Crottaz, P. Schobinger-Papamantellos, E. Suard, C. Ritter, S. Gentil, J.-P. Rivera and H. Schmid; Ferroelectrics \textbf{204}, 45-55 (1997); W. Sch\"{a}fer, G. Will, Physica Status Solidi (a), \textbf{28}, 211–215 (1975).\label{Mn3B7O13I}
\bibitem{Ni3B7O13Br} S. Y. Mao, F. Kubel, H. Schmid, P. Schobinger, P. Fischer, Ferroelectrics \textbf{146}, 81-97 (1993).\label{Ni3B7O13Br}
\bibitem{Co3B7O13Br} P. Schobinger-Papamantellos, P. Fischer, F. Kubel, H. Schmid; Ferroelectrics \textbf{162}, 93-101 (1994).\label{Co3B7O13Br}
\bibitem{Ni3B7O13Cl} Z.-G. Ye, P. Schobinger-Papamantellos, S. Y. Mao, F. Kubel, C. Ritter, E. Suard, M. Sato, H. Schmid, Ferroelectrics \textbf{204}, 83-95 (1997).\label{Ni3B7O13Cl}
\bibitem{Fe3B7O13OH} I. Nomoto, H. Sato, T. Fukui, Y. Narumi, K. Kindo, S. Nakamura, Y. Tsunoda, Journal of the Physical Society of Japan \textbf{80} 014801 (2011).\label{{Fe3B7O13OH}}
\bibitem{frustration} A. P. Ramirez, Annu. Rev. Mater. Sci. \textbf{24}, 453 (1994).\label{frustration}
\bibitem{fullprof} J. Rodriguez-Carvajal, Physica B \textbf{192}, 55 (1993) \label{fullprof}
\bibitem{lehmann} M. S. Lehmann, W. Kuhs, G. J. McIntyre, C. Wilkinson, J. Allibon; Journal of Applied Crystallography, \textbf{22}, 562-568 (1989).\label{lehmann}
\bibitem{Wilkinson} C. Wilkinson, H. W. Khamis, R. F. D. Stansfield, G. J. McIntyre, Journal of Applied Crystallography, \textbf{21}, 471-478 (1988).\label{Wilkinson}
\bibitem{eosphorite} G. D. Gatta, G. N\'{e}nert, P. Vignola, American Mineralogist, Volume \textbf{98}, pages 1297-1301 (2013).\label{eosphorite}
\bibitem{VESTA} K. Momma and F. Izumi, J. Appl. Crystallogr., \textbf{44}, 1272-1276 (2011).\label{VESTA}
\bibitem{BasiReps} J. Rodr\'{i}guez-Carvajal; BasIreps: A program for calculating irreducible representations of space groups and basis functions for axial and polar vector properties (see http://wwwold.ill.fr/dif/Soft/fp/php/downloads.html).\label{BasiReps}
\bibitem{Seitz} Seitz, F. (1934). Z. Kristallogr. 88, 433–459, Seitz, F. (1935a). Z. Kristallogr. 90, 289–313, Seitz, F. (1935b). Z. Kristallogr. 91, 336–366, Seitz, F. (1936). Z. Kristallogr. 94, 100–130; D. B. Litvin and V. Kopsk\'{y} Acta Cryst. (2011). A67, 415-418.\label{Seitz}
\bibitem{Co3B7O13Cl} M. Senthil Kumar, J.-P. Rivera, Z.-G. Ye, S. D. Gentil, H. Schmid, Ferroelectrics, \textbf{204}, 57-71 (1997).\label{Co3B7O13Cl}
\bibitem{Fe3B7O13X} D. Andreica , J.-P. Rivera , S. Gentil , Z.-G. Ye , M. Senthil Kumar, H. Schmid, Ferroelectrics, \textbf{204}, 73-81, (1997).\label{Fe3B7O13X}
\end{references}
\end{document}